\documentclass[a4paper,10pt]{article}

\usepackage{epsfig}
\usepackage{dcolumn}   % needed for some tables
\usepackage{bm}        % for math
\usepackage{amssymb}
\begin{document}

\begin{center} 
 
{\bf Hagedorn States and Thermalization}

J. Noronha-Hostler$^{1}$$^{2}$$^{3}$
C. Greiner$^{2}$

\vspace{0.5cm}
  $^{1)}$ {\it Frankfurt Institute for Advanced Studies (FIAS),} \\
          {\it Frankfurt am Main, Germany} \\
  $^{2)}$ {\it Institut f\"ur Theoretische Physik, Johann Wolfgang
Goethe--Universit\"at,} \\
          {\it Frankfurt am Main, Germany} \\
  $^{3)}$ {\it Helmholtz Research School,} \\
          {\it Frankfurt am Main, Germany} \\
\end{center}

\begin{abstract}
In recent years Hagedorn states have been used to explain the physics close to the critical temperature within a hadron gas.  Because of their large decay widths these massive resonances  lower $\eta/s$ to near the AdS/CFT limit within the hadron gas phase.  
A comparison of the Hagedorn model to
recent lattice results is made and it is found that for both Tc =176 MeV and Tc=196 MeV, the hadrons
can reach chemical equilibrium almost immediately, well before the chemical freeze-out
temperatures found in thermal fits for a hadron gas without Hagedorn states.  In this paper we also observe the effects of Hagedorn States on the $K^+/\pi^+$ horn seen at AGS, SPS, and RHIC. 
\end{abstract}

\date{\today}

\section{Introduction}

As two heavy ions collide color neutral clusters are formed within which the number of particles per cluster increase.  The clusters become so dense and begin to overlap such that it impossible to distinguish quarks from one cluster from that in another i.e. a percolation transition.  The critical density for this is about 
$\epsilon\approx 1\; GeV/fm^3$.
Following the phase transition into Quark Gluon Plasma the interactions are dominated by quarks and gluons.  Through gluon fusions, strange quarks can easily be reproduced. Eventually the QGP cools back into hadrons where the particle yields and ratios are measured.  

If one only considers binary collisions, which react too slowly for strange particles to reach chemical equilibrium within the hadron gas phase, then it is clear that strange particle yields can only be explained through gluon fusion within QGP \cite{Koch:1986ud} and that the hadrons must exist QGP already in full chemical equilibrium \cite{Stock:1999hm}. 
However, multi-mesonic collisions 
$n\pi\leftrightarrow X\bar{X}$
have been demonstrated to reach chemical equilibration
for various (strange) antibaryons quickly at SPS  \cite{Rapp:2000gy,Greiner}, although they are still not enough to explain the particle yields of exotic antibaryons at the higher energies at RHIC  \cite{Kapusta,Huovinen:2003sa}.  
In order to circumvent such
longer time scales $\sim 10 $ fm/c for a situation of
a nearly baryon-free system with nearly as much antibaryons as baryons, it was then suggested
by Braun-Munzinger, Stachel and Wetterich \cite{BSW} that near Tc there exists
an extra large particle density overpopulated with pions
and kaons, which then drive the baryons/anti-baryons into
equilibrium by exactly such multi-mesonic collisions. But it is not clear how and why this overpopulation of pions and kaons
should appear, and how the subsequent population
of (anti-)baryons would follow in accordance with a standard statistical
hadron model: According to the mass action law the overpopulated
matter of pions will result in an overpopulation of 
(anti-)baryons. For such a large number of
(anti-)baryons it is difficult to get rid of them quickly enough in order to
reach standard hadron equilibrium values before the chemical freeze-out \cite{Greiner:2004vm}.

Rather, understanding the rapid chemical equilibration is possible using Hagedorn states, heavy resonances that drive similar and more multi-hadronic reactions close to $T_c$, as shown in \cite{Greiner:2004vm,NoronhaHostler:2007fg,NoronhaHostler:2007jf,NoronhaHostler:2009hp,long}.  
Close to $T_c$ the matter is then a strongly interacting mixture
of standard hadrons and such resonances. Using the Hagedorn states as potential and highly unstable catalysts, the standard hadrons can be populated reactions:
\begin{eqnarray}\label{eqn:decay}
n\pi&\leftrightarrow &HS\leftrightarrow n^{\prime}\pi+X\bar{X}
\end{eqnarray} 
where $X\bar{X}$ can be substituted with  $p\bar{p}$ , $K\bar{K}$ , $\Lambda\bar{\Lambda}$, or $\Omega\bar{\Omega}$.   
The large masses of the decaying Hagedorn states open up the
phase space for multi-particle decays. 

In this note we will compare the particle ratios obtained by using reactions driven by Hagedorn states and those of the experiments at RHIC.  We find that both strange and non-strange particles match the experimental data well within the error bars.  Furthermore, we are able to make estimates for the chemical equilibration time and find that they are very short, which implies that the hadrons can easily reach chemical equilibrium within an expanding, hadronic fireball and that hadrons do not need to be ``born" into chemical equilibrium \cite{NoronhaHostler:2007jf,NoronhaHostler:2009hp,long}.
Hagedorn states thus provide a microscopic basis for understanding hadronization
of deconfined matter to all hadronic particles.

Before starting with the details, we emphasize that Hagedorn states have become
quite popular to understand the physics of strongly interacting matter close to the critical temperature:
Hagedorn states have been shown to contribute to the physical description of a hadron gas close to $T_c$.  The inclusion of Hagedorn states leads to a low $\eta/s$ in the hadron gas phase \cite{NoronhaHostler:2008ju}, which nears the string theory bound $\eta/s=1/(4\pi)$ . Calculations of the trace anomaly including Hagedorn states also fits recent lattice results well and correctly describe the minimum of  the speed of sound squared, $c_s^2,$ near the phase transition found on the lattice \cite{NoronhaHostler:2008ju}. Estimates for the bulk viscosity including Hagedorn states in the hadron gas phase indicate that the bulk viscosity, $\zeta/s$, increases near $T_c$ \cite{NoronhaHostler:2008ju}.
We also remark here that Hagedorn states can also explain the phase transition
above the critical temperature and, depending on the
intrinsic parameters, the order of the phase transition
\cite{Zakout:2006zj}.

Because of the success of thermal models when fitting experimental data at RHIC, SPS, and AGS \cite{thermalmodels,Rafelski:2008av,Schenke:2003mj,RHIC,Andronic:2005yp,Becattini:2005xt,Manninen:2008mg}, a study was done on the effect of adding in the influence of Hagedorn states to the thermal models at RHIC energies \cite{NoronhaHostler:2009tz,NoronhaHostler:2010yc}. It was found that not only does the addition of Hagedorn states improve the $\chi^2$ of the fit but that the addition of Hagedorn states increases slightly the chemical freeze-out temperature.
Due to the success of the implementation of Hagedorn states into other aspects of hadronic gas physics, we have decided to investigate the possible effects that Hagedorn states would have on the horn seen in the $K^+/\pi^+$ ratio \cite{Gazdzicki:2004ef,:2007fe}.  A recent study has investigated the effects by adding in the decay of Hagedorn states into pions \cite{Andronic:2008gu}.  In this proceedings we will include the effects of Hagedorn states on all particles, not just the pions, in order to observe the horn. We find that the Hagedorn states do not contribute significantly to the horn. 

\section{Setup}

The basis of the Hagedorn spectrum is that there is an exponential mass increase along with a prefactor i.e. the mass spectrum has the form: $f(m)\approx\exp^{m/T_H}$ \cite{Hagedorn:1968jf}. The exponential mass spectrum drives open the phase space, which allows for multi-mesonic decays to dominate close to $T_c$ (we assume $T_H\approx T_c$).    We use the form
\begin{equation}\label{eqn:fitrho}
    \rho=\int_{M_{0}}^{M}\frac{A}{\left[m^2 +m_{r}^2\right]^{\frac{5}{4}}}e^{\frac{m}{T_{H}}}dm.
\end{equation}
where $M_{0}=2$ GeV and $m_{r}^2=0.5$ GeV.  We consider  two different different lattice results for $T_c$: $T_c=196$ MeV \cite{Cheng:2007jq,Bazavov:2009zn} (the corresponding fit to the trace anomaly is then $A=0.5 GeV^{3/2}$, $M=12$ GeV, and $B=\left(340 MeV\right)^4$), which uses an almost physical pion mass, and $T_c=176$ MeV \cite{zodor} (the corresponding fit to the energy density leads to $A=0.1 GeV^{3/2}$, $M=12$ GeV, and $B=\left(300 MeV\right)^4$). Both are shown and discussed in \cite{long}. Furthermore, we need to take into account the repulsive interactions and, thus,  use volume corrections  \cite{long,NoronhaHostler:2008ju,Kapusta:1982qd},
which ensure that the our model is thermodynamically consistent. Note that $B$ is a free parameter  based upon the idea of the MIT bag constant.  

We need to consider the back reactions of multiple particles combining to form a Hagedorn state in order to preserve detailed balance.  
The rate equations for the Hagedorn resonances $N_{i}$, pions $N_{\pi}$, and the $X\bar{X}$ pair $N_{X\bar{X}}$, respectively, are given by
\begin{eqnarray}\label{eqn:setpiHSBB}
\noindent\dot{N}_{i}&=&\Gamma_{i,\pi}\left[N_{i}^{eq}\sum_{n} B_{i,n}
\left(\frac{N_{\pi}}{N_{\pi}^{eq}}\right)^{n}-N_{i}\right]+\Gamma_{i,X\bar{X}}\left[ N_{i}^{eq}
\left(\frac{N_{\pi}}{N_{\pi}^{eq}}\right)^{\langle n_{i,x}\rangle} \left(\frac{N_{X\bar{X}}}{N_{X\bar{X}}^{eq}}\right)^2 -N_{i}\right]\nonumber\\
\dot{N}_{\pi }&=&\sum_{i} \Gamma_{i,\pi}  \left[N_{i}\langle n_{i}\rangle-N_{i}^{eq}\sum_{n}
B_{i, n}n\left(\frac{N_{\pi}}{N_{\pi}^{eq}}\right)^{n} \right]\nonumber\\
&+&\sum_{i} \Gamma_{i,X\bar{X}} \langle n_{i,x}\rangle\left[N_{i}-
N_{i}^{eq}
\left(\frac{N_{\pi}}{N_{\pi}^{eq}}\right)^{\langle n_{i,x}\rangle} \left(\frac{N_{X\bar{X}}}{N_{X\bar{X}}^{eq}}\right)^2\right]  \nonumber\\
\dot{N}_{X\bar{X}}&=&\sum_{i}\Gamma_{i,X\bar{X}}\left[ N_{i}- N_{i}^{eq}\left(\frac{N_{\pi}}{N_{\pi}^{eq}}\right)^{\langle n_{i,x}\rangle} \left(\frac{N_{X\bar{X}}}{N_{X\bar{X}}^{eq}}\right)^2\right].
\end{eqnarray}
The decay widths for the $i^{th}$ resonance are $\Gamma_{i,\pi}$ and $\Gamma_{i,X\bar{X}}$, the branching ratio is $B_{i,n}$ (see below), and the average number of pions that each resonance will decay into is $\langle n_{i}\rangle$.  The equilibrium values $N^{eq}$ are both temperature and chemical potential dependent.  However, here we set $\mu_b=0$.
Additionally, a discrete spectrum of Hagedorn states is considered, which is separated into mass bins of 100 MeV. 

The branching ratios, $B_{i,n}$, are the probability that the $i^{th}$ Hagedorn state will decay into $n$ pions where $\sum_{n}B_{i,n}=1$ must always hold. 
We assume the branching ratios follow a Gaussian distribution for the reaction $HS\leftrightarrow n\pi$ 
\begin{equation}
B_{i, n}\approx
\frac{1}{\sigma_{i}\sqrt{2\pi}}e^{-\frac{(n-\langle n_{i}\rangle)^{2}}{2\sigma_{i} ^{2}}},
\end{equation}
which has its peak centered at $\langle n_{i}\rangle\approx 3 - 34$ and the width of the distribution is $\sigma_{i}^2\approx0.8 - 510$ (see \cite{long}). For the average number of pions when a $X\bar{X}$ pair is present, we again refer to the micro-canonical model in \cite{Greiner:2004vm,Liu} and find
\begin{equation}\label{eqn:nfit}
    \langle n_{i,x}\rangle=\left(\frac{2.7}{1.9}\right)\left(0.3+0.4m_i\right)\approx 2-7.
\end{equation}
where $m_i$ is in GeV. In this paper we do not consider a distribution but rather only the average number of pions when a $X\bar{X}$ pair is present.  We  assume that $\langle n_{i,x}\rangle=\langle n_{i,p}\rangle=\langle n_{i,k}\rangle=\langle n_{i,\Lambda}\rangle=\langle n_{i,\Omega}\rangle$ for when a kaon anti-kaon pair, $\Lambda\bar{\Lambda}$, or  $\Omega\bar{\Omega}$ pair is present. 

The decays widths are defined as follows (see \cite{long}):
\begin{eqnarray}\label{decaywidth}
\Gamma_{i}&=&0.15m_{i}-0.0584=250\;\mathrm{MeV\;to}\;1800 \;\mathrm{MeV}\nonumber\\
\Gamma_{i,X\bar{X}}&=&\langle X_i\rangle \Gamma_{i}\nonumber\\
\Gamma_{i,\pi}&=&\Gamma_{i}-\Gamma_{i,X\bar{X}}.
\end{eqnarray}
$\Gamma_{i}$ is a linear fit extrapolated  from the data in \cite{Eidelman:2004wy}. It is then separated into two parts, one for the reaction $HS\leftrightarrow n\pi$ i.e. $\Gamma_{i,\pi}$ and one for the reaction $HS\leftrightarrow n\pi+X\bar{X}$ i.e.  $\Gamma_{i,X\bar{X}}$.  The decay width $\Gamma_{i,X\bar{X}}$ is found my multiplying $\langle X_i\rangle$, which is the average X that a Hagedorn state of mass $m$ will decay into, that is found from both microcanonical \cite{Greiner:2004vm,Liu} and canonical models  \cite{long}. 
The large masses open up the phase space for such more special multi-particle decays.
A detailed explanation is found in \cite{long}.

The equilibrium values are found using a statistical model \cite{StatModel}, which includes 104 light or strange particles from the the PDG \cite{Eidelman:2004wy}. Throughout this paper our initial conditions are the various fugacities at $t_0$ (at the point of the phase transition into the hadron gas phase)
$\alpha\equiv\lambda_{\pi}(t_0) \, , \, \beta_{i}\equiv\lambda_{i}(t_0) \, , \mbox{and}\,   \phi\equiv\lambda_{X\bar{X}}(t_0)$
which are chosen by holding the contribution to the total entropy from the Hagedorn states and pions constant i.e. 
$s_{Had}(T_{0},\alpha)V(t_{0})+s_{HS}(T_{0},\beta_{i})V(t_{0})=s_{Had+HS}(T_{0})V(t_{0})=const$
and the corresponding initial condition configurations we choose later can be seen in Tab.\ \ref{tab:IC} (for further discussion see \cite{long}).In our model we do not just consider the direct number of hadrons but also the indirect number that comes from other resonances.  For example, for pions we consider also the contribution from resonances such as $\rho$'s, $\omega$'s etc.  The number of indirect hadrons can be calculated from the branching ratios for each individual species in the particle data book \cite{Eidelman:2004wy}. Moreover, there is also a contribution from the Hagedorn states to the total number of pions, kaons, and so on as described in \cite{NoronhaHostler:2007jf,long}.  Thus, the total number of ``effective" pions can be described by
\begin{eqnarray}\label{eqn:effpi}
\tilde{N}_{\pi}&=&N_{\pi}+\sum_{i}N_{i}\langle n_{i}\rangle
\end{eqnarray}
whereas the total number of ``effective" p's, $K$'s, $\Lambda$'s,  etc. (generalized as $X$) can be described by
\begin{eqnarray}\label{eqn:effbbkk}
\tilde{N}_{X}&=&N_{X}+\sum_{i}N_{i}\langle X_{i}\rangle.
\end{eqnarray}
Because the Hagedorn states are relevant only near $T_c$ , the contribution of the Hagedorn states to the total particles numbers if only effected close to $T_c$.

\section{Results: Expansion}

In order to include the cooling of the fireball we need to find a relationship between the temperature and the time, i.e., $T(t)$.  To do this we apply a Bjorken expansion for which the total entropy is held constant
\begin{equation}\label{eqn:constrain}
\mathrm{const.}=s(T)V(t)\sim\frac{S_{\pi}}{N_{\pi}}\int \frac{dN_{\pi}}{dy} dy.
\end{equation}
where $s(T)$ is the entropy density of the hadron gas with volume corrections.  The total number of pions in the $5\%$ most central collisions, $\frac{dN_{\pi}}{dy}$, can be found from experimental
results in \cite{Bearden:2004yx}.  Thus, our total pion number is
$\sum_{i}N_{\pi^{i}}=\int_{-0.5}^{0.5} \frac{dN_{\pi}}{dy} dy=874$.
While for a gas of non-interacting Bose gas of massless pions $S_{\pi}/N_{\pi}=3.6$, we do have a mass for a our pions, so we must adjust $S_{\pi}/N_{\pi}$ accordingly.  In \cite{Greiner:1993jn} it was shown that when the pions have a mass the ratio changes and, therefore, the entropy per pion is close to $S_{\pi}/N_{\pi}\approx5.5$, which is what we use here.

The effective volume at mid-rapidity can be parametrized as a function of time.  We do this by using a Bjorken expansion and including accelerating radial flow.  
The volume term is then
\begin{equation}\label{eqn:bjorken}
V(t)=\pi\;ct\left(r_{0}+v_{0}(t-t_{0})+\frac{1}{2}a_{0}(t-t_{0})^2 \right)^2
\end{equation}
where the initial radius is $r_{0}(t_0)=7.1$ fm for $T_H=196$ and the corresponding $t_{0}^{(196)}\approx2 fm/c$. For $T_H=176$ we allow for a longer expansion before the hadron gas phase is reached and, thus, calculate the appropriate $t_0^{(176)}$ from the expansion starting at $T_H=196$, which is $t_0^{(176)}\approx 4 fm/c$. We use $v_0=0.5 $ and $a_0=0.025 $ (see \cite{long}).

Because the volume expansion depends on the entropy and the Hagedorn resonances contribute strongly to the entropy only close to the critical temperature (see \cite{long}),  the effects of the Hagedorn states must be taken into account with calculating the total particle yields otherwise the yields do not increase with the temperature (see \cite{long} for further discussion).  This is precisely what is done in Eqs. (\ref{eqn:effpi}) and (\ref{eqn:effbbkk})
because Hagedorn states also contribute strongly to the $\pi$'s and $X\bar{X}$ pairs close to $T_c$.

\begin{table}
\begin{center}
 \begin{tabular}{|c|c|c|c|}
 \hline
 & & & \\
   & $\alpha=\lambda_{\pi}(t_0)$ & $\beta_{i}=\lambda_i(t_0)$ & $\phi=\lambda_{X\bar{X}}(t_0)$ \\
    & & & \\
 \hline
$IC_1$ & 1 & 1 & 0 \\
$IC_2$ & 1 & 1 & 0.5 \\
$IC_3$ & 1.1 & 0.5 & 0 \\
$IC_4$ & 0.95 & 1.2 & 0 \\
 \hline
 \end{tabular}
 \end{center}
 \caption{Initial condition configurations.}\label{tab:IC}
 \end{table}
Along with the expansion we also must solve these rate equations, Eq. (\ref{eqn:setpiHSBB}), numerically.  We start with various initial conditions as seen in table II and the initial temperature is the respective critical temperature and we stop at $T=110$ MeV. 
\begin{figure}
\centering
\epsfig{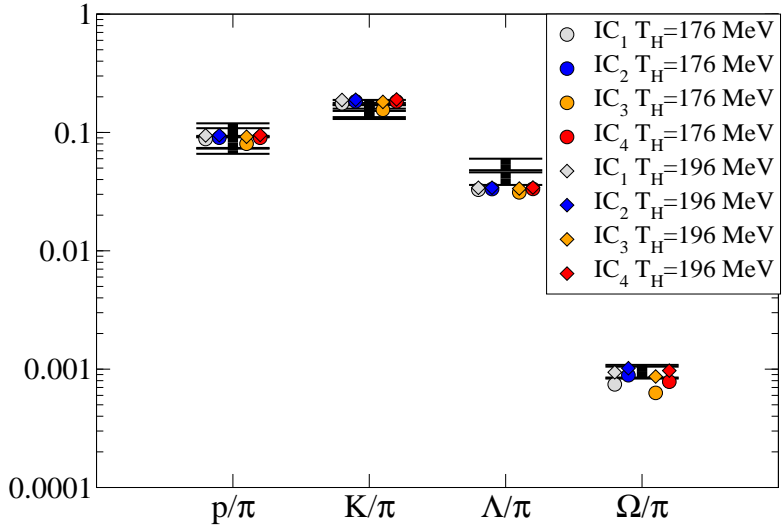}
\caption{Plot of the various ratios including all initial conditions defined in Tab.\ \ref{tab:IC}.  The points show the ratios at $T=110$ MeV for the various initial conditions (circles are for $T_H=176$ MeV and diamonds are for $T_H=196$ MeV).   The experimental results for STAR and PHENIX are shown by the black error bars.} \label{fig:summary}
\end{figure}
A summary graph of all our results is shown in Fig.\ \ref{fig:summary}.  The black error bars cover the range of error for the experimental data points from STAR and PHENIX. 
 The points show the range in values for the initial conditions at a final
expansion point with a temperature $T=110$ MeV. 
We see in our graph that our freezeout results match the experiments well and the initial conditions have little effect on the ratios, which implies that information from the QGP regarding multiplicities is washed out due to the rapid dynamics of Hagedorn states.  
A smaller $\beta_i$ slows the equilibrium time slightly.  However, as seen in Fig.\ \ref{fig:summary} it still fits within the experimental values. Further discussion of the effects of our chosen decay widths can be found in \cite{long}, as well as individual results for that ratios within an expanding firebal.   Furthermore, in \cite{NoronhaHostler:2007jf} we showed the the initial condition play almost no roll whatsoever in  $K/\pi^{+}$ and  $(B+\bar{B})/\pi^{+}$.  Thus, strengthening our argument that the dynamics are washed out following the QGP.

\section{$K^+/\pi^+$ Horn}\label{model}

The $K^+/\pi^+$ ratio was first discussed in \cite{Gazdzicki:2004ef,:2007fe} and has yet to be accurately explained using thermal models.  However, it has been suggested that Hagedorn states could possibly be the explanation for the horn \cite{Andronic:2008gu}.  Using the $T_{ch}$ and $\mu_b$ given in \cite{Andronic:2005yp} we calculate the strange chemical potential, $\mu_s$, with the conservation of strangeness
\begin{eqnarray}\label{eqn:cons}
\frac{\sum_i n_i S_i}{\sum_i n_i B_i}&=&0,
\end{eqnarray}
we are then able to calculate the corresponding $K^+/\pi^+$ at each experimental data point. At present we do not conserve charge (or rather isospin), however, we are currently working on a model that includes the electrical charge. We used data from RHIC, SPS, and AGS.  The citations for the experimental data can all be found in \cite{RHIC,Andronic:2005yp}.  

\begin{figure}
\centering
\epsfig{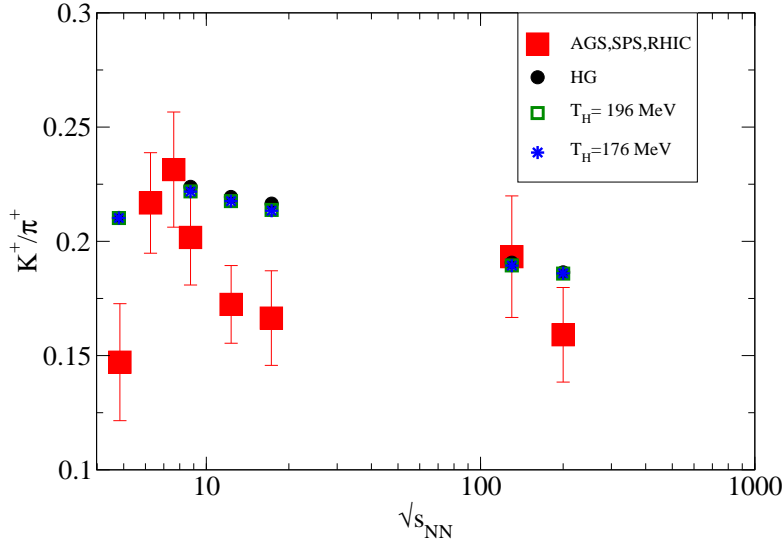}
\caption{Thermal model results for the $K^+/\pi^+$ ratio at various energies without Hagedorn states and with using two different Hagedorn temperatures: $T_H=176$ MeV and $T_H=196$ MeV.  A comparison is shown to the data from AGS, SPS, and RHIC.  } \label{fig:horn}
\end{figure} 

In Fig.\ \ref{fig:horn} we have plotted the $K^+/\pi^+$ ratio versus $\sqrt{s_{NN}}$.  The experimental data points are shown with the error bars while our pure hadron gas is a dot.  The two thermal model results with Hagedorn states are the square and star, which represent $T=196$ MeV and $T_H=176$ MeV, respectively.  One can clearly see from the graph that there is almost no difference between the three different results.  This is not surprising because at lower beam energies, the chemical freezeout temperature is also lower.  Around the peak of the horn the chemical freeze-out temperature range from $T=124-160$ MeV, which means that most of the Hagedorn states have already died out even from the lower critical temperature of $T_H=176$ MeV and have long since died out from $T_H=196$ MeV. One can clearly see from the effects of the Hagedorn states on the total particle yields in Fig.\ 14 and Fig.\ 15 in \cite{long} that at $T=160$ MeV the Hagedorn states have almost no effect on the particle yields regardless of the $T_H$. If one were to lower the critical temperature closer to $T\approx 160$ MeV then there might be a stronger influence of the Hagedorn states on the horn.  In recent lattice calculations it has been suggested that the critical temperature might be lower \cite{Borsanyi:2010cj}, which we plan to look at in the future. Additionally, in our upcoming paper we will refit the results using a thermal model that conserves baryon number, charge, and strangeness.

\section{Conclusion}
\label{con}

In this paper we discussed the effects of Hagedorn states on the $K^+/\pi+$ horn, which were found to be negligible.  Because the $K^+/\pi+$ horn is measured at lower beam energies than RHIC (and, hence, the typical temperatures are significantly below $T_H$  i.e. $T_{ch}<160$ MeV at the peak of the horn), it is not surprising that Hagedorn states do not play a role because Hagedorn states  are highly suppressed far from the critical temperature. It is interesting to note that if recent lattice calculations are correct that exhibit a lower critical temperature region \cite{Borsanyi:2010cj} then possibly the Hagedorn states could effect the $K^+/\pi+$ horn, which we will attempt to study in a future paper.  An attempt to look at Hagedorn states within this new lattice framework was shown in \cite{Majumder:2010ik}.  However, the repulsive interactions were not taken into account.  As a future project we will create a thermal model that conserves baryon number, strangeness, and charge that looks specifically at the effects of Hagedorn states on thermal fits for energy ranges at AGS, SPS, and RHIC.  

The Hagedorn states provide a mechanism for quick chemical equilibration times. Our model gives chemical equilibration times on the order of $\Delta \tau\approx 1-3\frac{fm}{c}$. Furthermore, the particle ratios obtained from decays of Hagedorn states match the experimental values at RHIC very well, which leads to the conclusion that hadrons do not need to be born in chemical equilibrium.    Rather a scenario of hadrons that reach chemical freeze-out shortly after the critical temperature due to multi-mesonic reactions driven by Hagedorn states, is entirely plausible.  We have shown that both strange ($\Lambda$'s and K's) and non-strange ($\pi$'s and $p$'s) hadrons can reach chemical equilibration by $T=160$ MeV. Thus, it would be interesting to implement Hagedorn states into a transport approach such as UrQMD \cite{URQMD}.  Such multi-quark droplets are clearly recognized when looking at effective models of hadronization like the chromodieletric model \cite{Martens:2004ad}. 
Moreover, even multi-strange baryons such as $\Omega$'s can reach chemical equilibrium in such a scenario. Our work indicates that the population and repopulation of potential
Hagedorn states close to phase boundary 
can be the key source for a dynamical understanding of generating and
chemically equilibrating the standard and measured hadrons. 
Hagedorn states thus can provide a microscopic basis for understanding hadronization
of deconfined matter.

\section{Acknowledgments}

This work was supported by the Helmholtz International Center
for FAIR within the framework of the LOEWE program (Landes-Offensive zur Entwicklung
Wissenschaftlich-¨okonomischer Exzellenz) launched by the State of Hesse.

%%%%%%%%%%%%%%%%%%%%%%%%%%%%%%%%%%%%%%%%%%%%%%%%%%%%%%%%%%%%%%%%%%%%%%%

%%%%%%%%%%%%%%%%%%%%%%%%%%%%%%%%%%%%%%%%%%%%%%%%%%%%%%%%%%%%%%%%%%%%%%%

\end{document}